\documentclass{appolb}
\usepackage{epsfig}
\usepackage{lineno} 

\renewcommand{\d}{\mathrm{d}}
\newcommand{\pt}{p_\mathrm{T}}
\newcommand{\mt}{m_\mathrm{T}}

\begin{document}
\title{Light neutral mesons production in p-A collisions at $\sqrt{s} = 27.5$~GeV with the NA60 Experiment
\thanks{\scriptsize{Presented at Strangeness in Quark Matter 2011, 18-24 Sept. 2011, Cracow, Poland}}
}
\author{Antonio Uras\footnote{\scriptsize{IPNL, Universit\'e de Lyon, Universit\'e Lyon 1, CNRS/IN2P3, Villeurbanne, France}} for the NA60 Collaboration}

\maketitle


\begin{abstract}
\noindent The NA60 experiment has studied low-mass muon pair production in proton-nucleus (p-A) collisions with a system of Be, Cu, In, W, Pb and~U targets using a 400 GeV/$c$ proton beam at the CERN SPS. Thanks to the collected data sample of 180\,000 low mass muon pairs, the most precise measurement currently available was performed for the electromagnetic transition form factors of the $\eta$ and $\omega$ mesons. The $\rho$ line shape was quantitatively investigated, and its effective temperature measured for the first time in elementary collisions. The transverse momentum spectra for the $\omega$ and $\phi$ mesons have been studied in the full $\pt$ range accessible, up to 2~GeV/$c$. The cross section ratios $\sigma_\rho/\sigma_\omega$ and $\sigma_\phi/\sigma_\omega$ have been considered in full~$\pt$ as a function of the size of the production target. The nuclear dependence of the production cross sections of the $\eta$, $\omega$ and $\phi$ mesons has finally been studied in terms of the power law $\sigma_\mathrm{pA} \propto \mathrm{A}^\alpha$, where the $\alpha$ parameter has been found to increase as a function of $\pt$.
\end{abstract}

\PACS{14.40.Be, 13.20.Jf, 25.75.-q} 
  
~\\
\noindent The study of the production of low mass vector and pseudoscalar mesons in p-A collisions offers several points of interest. First, it allows one to perform measurements of particle properties in an environment of cold nuclear matter, as in the case of the line shape of the $\rho$ meson and the electromagnetic transition form factors of the $\eta$ and $\omega$ mesons. In a second place, p-A data represent a natural baseline for the heavy-ion observations, providing a reference in an environment of cold nuclear matter for several observables such as strangeness production as a function of the size of the nucleus. Finally, the study of the nuclear dependence of particle properties, as the transverse momentum spectra and the production cross sections, is an effective tool to understand the dynamics of soft hadron interactions. In order to address these items, the NA60 experiment has studied low-mass muon pair production in p-A collisions at the CERN SPS, with a system of Be, Cu, In, W, Pb and~U targets, collecting $\sim 180\,000$ low mass muon pairs at $\sqrt{s} = 27.5$~GeV.

\noindent A description of the NA60 apparatus can be found for example in~\cite{apparatus}. The produced dimuons are identified and measured by the muon spectrometer placed after a hadron absorber. The latter allows one to select the highly rare dimuon events but induces at the same time multiple scattering and energy loss on the muons, thus degrading the mass resolution of the measurement made in the spectrometer. To overcome this problem, NA60 already measures the muons before the absorber, with a silicon pixel spectrometer. The final sample of dimuons retained for the analysis is defined according to the quality of the matching between the reconstructed tracks in the two spectrometers: this leads to a total sample of $\sim 180\,000$ dimuons. When the identification of the production target is requested, however, stricter cuts must be applied, reducing the total sample to $\sim 80\,000$ dimuons. The small component of the combinatorial background ($\pi$ and $K$ decays) is estimated via an event mixing technique and subtracted from the real data, with an overall precision of a few percent. More details can be found in~\cite{HP2010}.

\begin{figure}[htbp] 
  \begin{center}
    \includegraphics[width=0.49\textwidth]{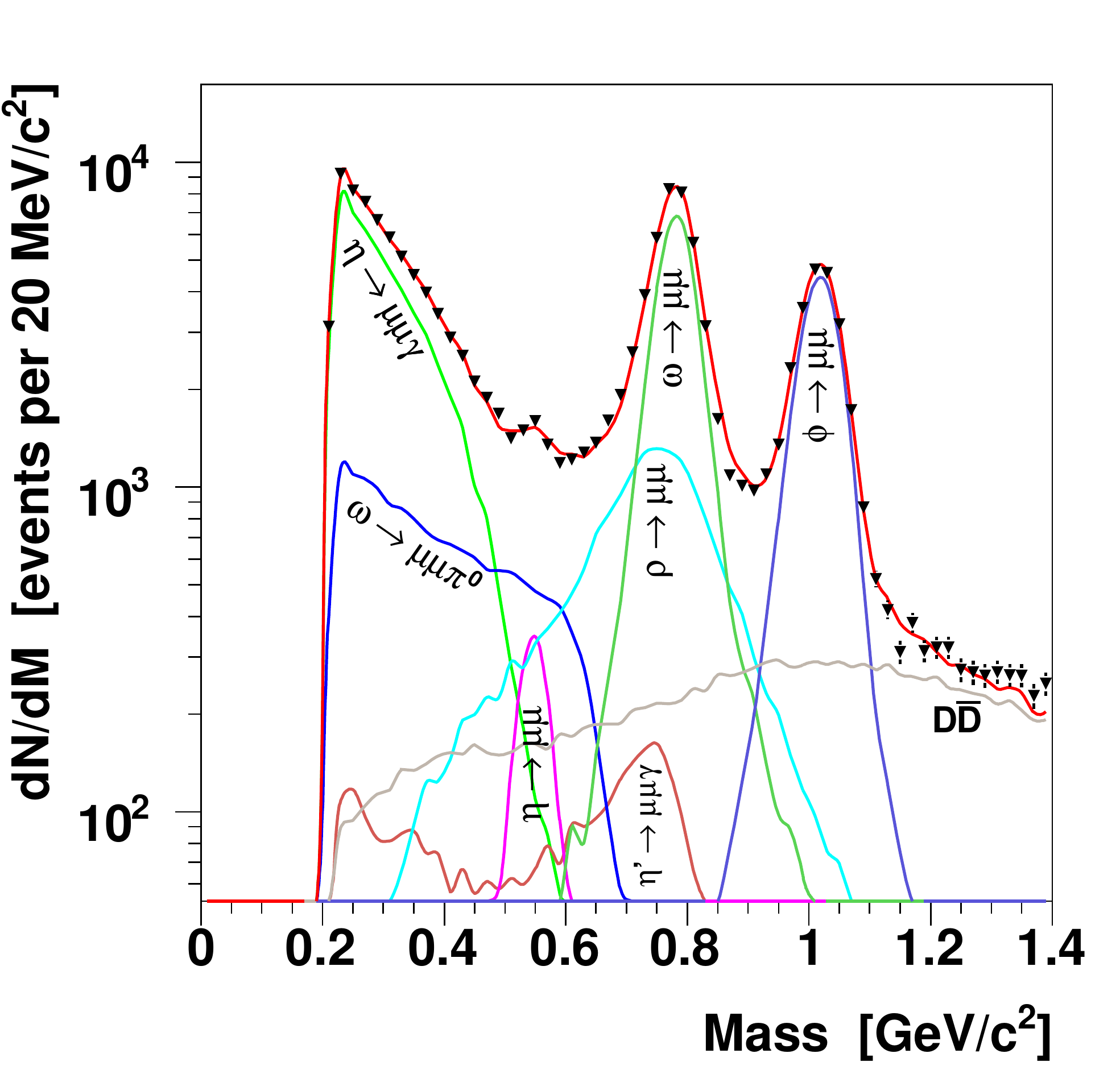}
    \includegraphics[width=0.49\textwidth]{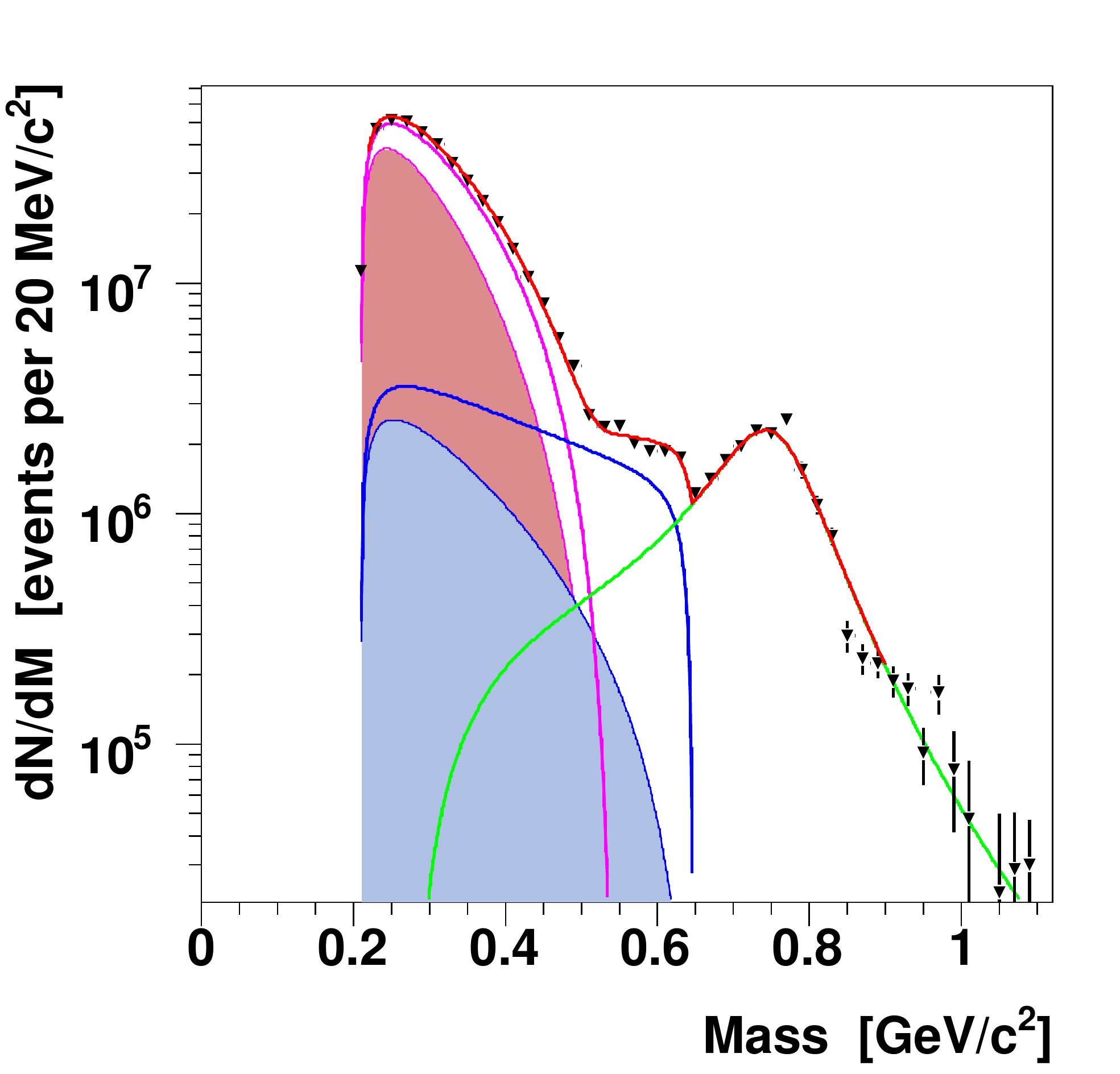}
  \end{center}
\vspace{-0.5cm}
\caption[\textwidth]{Left panel: fit on the target-integrated raw mass spectrum. Right panel: fit on the acceptance corrected mass spectrum relative to the processes $\eta\to\mu^+\mu^-\gamma$, $\omega\to\mu^+\mu^-\pi^0$ and $\rho\to\mu^+\mu^-$. The shaded areas indicate the Kroll-Wada expectations for point-like particles, defined by QED~\cite{KW}.}
\label{fig:mass_spectra}
\end{figure}

~\\
\noindent As it can be seen from the left panel of \figurename~\ref{fig:mass_spectra}, the dimuon mass spectrum is well described by the superposition of the two-body and Dalitz decays of the light neutral mesons $\eta$, $\rho$, $\omega$, $\eta'$ and $\phi$, with an additional open charm component. Considering the target-integrated data sample, the electromagnetic transition form factors of the $\eta$ and $\omega$ mesons have been studied through the Dalitz decays $\eta \to \mu^+ \mu^- \gamma$ and $\omega \to \mu^+ \mu^- \pi^0$. These form factors are usually expressed in terms of the pole parametrization $|F|^2 = (1 - M^2/\Lambda^2)^{-2}$, with the Vector Meson Dominance (VMD) model giving theoretical predictions both for $\Lambda_{\eta}^{-2}$ and $\Lambda_{\omega}^{-2}$, which can now be tested with the NA60 p-A data. While the details of the analysis can be found elsewhere~\cite{Uras:HQ2010}, we report here the final values, extracted from the fit on the corrected mass spectrum shown in the right panel of \figurename~\ref{fig:mass_spectra}, currently the most precise measurements available these parameters: $\mathrm{\Lambda}_\eta^{-2} = 1.951\ \pm\ 0.059$~(stat.) $\pm\ 0.042$~(syst.) and $\mathrm{\Lambda}_\omega^{-2} = 2.240\ \pm\ 0.024$~(stat.) $\pm\ 0.028$~(syst.), both expressed in (GeV/$c^2$)$^{-2}$. They are in perfect agreement with the values obtained by the analysis on the NA60 peripheral In-In data~\cite{na60} and the Lepton-G results~\cite{landsberg}, confirming the agreement with the VMD model for the $\eta$ meson and the discrepancy in the case of the $\omega$ meson. An improved description of the $\omega$ form factor, given by an alternative theoretical approach recently appeared in~\cite{Leupold}, still underestimates the part close to the kinematical cut-off. The same analysis which lead to the measurement of the form factors, also allowed us to isolate and study the line shape of the $\rho$ meson, which has been parametrized with the function 
\begin{displaymath}
{\textstyle \frac{\d N}{\d M} \propto \frac{\sqrt{1-4m^2_\mu/M^2}\left( 1+2m^2_\mu/M^2\right)\left( 1-4m^2_\pi/M^2\right) ^{3/2}}
{\left( m^2_{\rho}-M^2\right)^2+m^2_\rho\Gamma^2_\rho(M)} \left(MT\right)^{3/2} e^{-\frac{M}{T}}}~.
\end{displaymath}
As it can be seen from the right panel of \figurename~\ref{fig:mass_spectra}, this parametrization is well suited to describe the $\rho$ line shape in cold nuclear matter, with the Boltzmann-like factor $\left(MT\right)^{3/2} e^{-\frac{M}{T}}$ playing a crucial role in shaping the low mass tail. The $T$ parameter is found to be $160\pm5$~(stat.) $\pm\:7$~(syst.)~MeV. This represents the first measurement of the effective temperature of the $\rho$ meson in cold nuclear matter: it is in agreement with the measurement in peripheral In-In~\cite{na60}, and consistent with the value of 170~MeV obtained by statistical model fits on particle ratios in p-p interactions.

\begin{figure}[htbp] 
  \begin{center}
    \vspace{-0.3cm}		
    \includegraphics[width=0.40\textwidth]{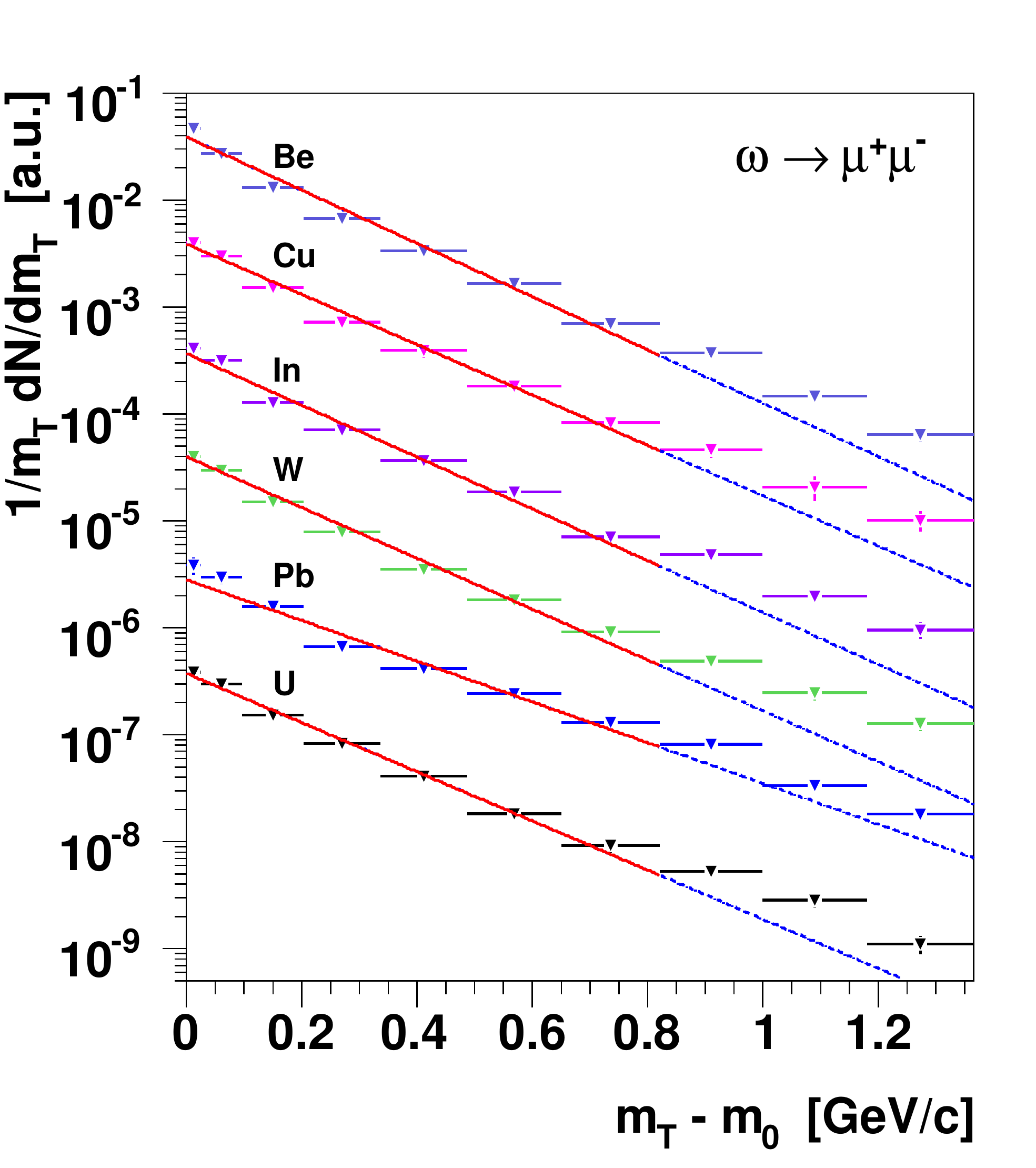}\hspace{0.05\textwidth}
    \includegraphics[width=0.40\textwidth]{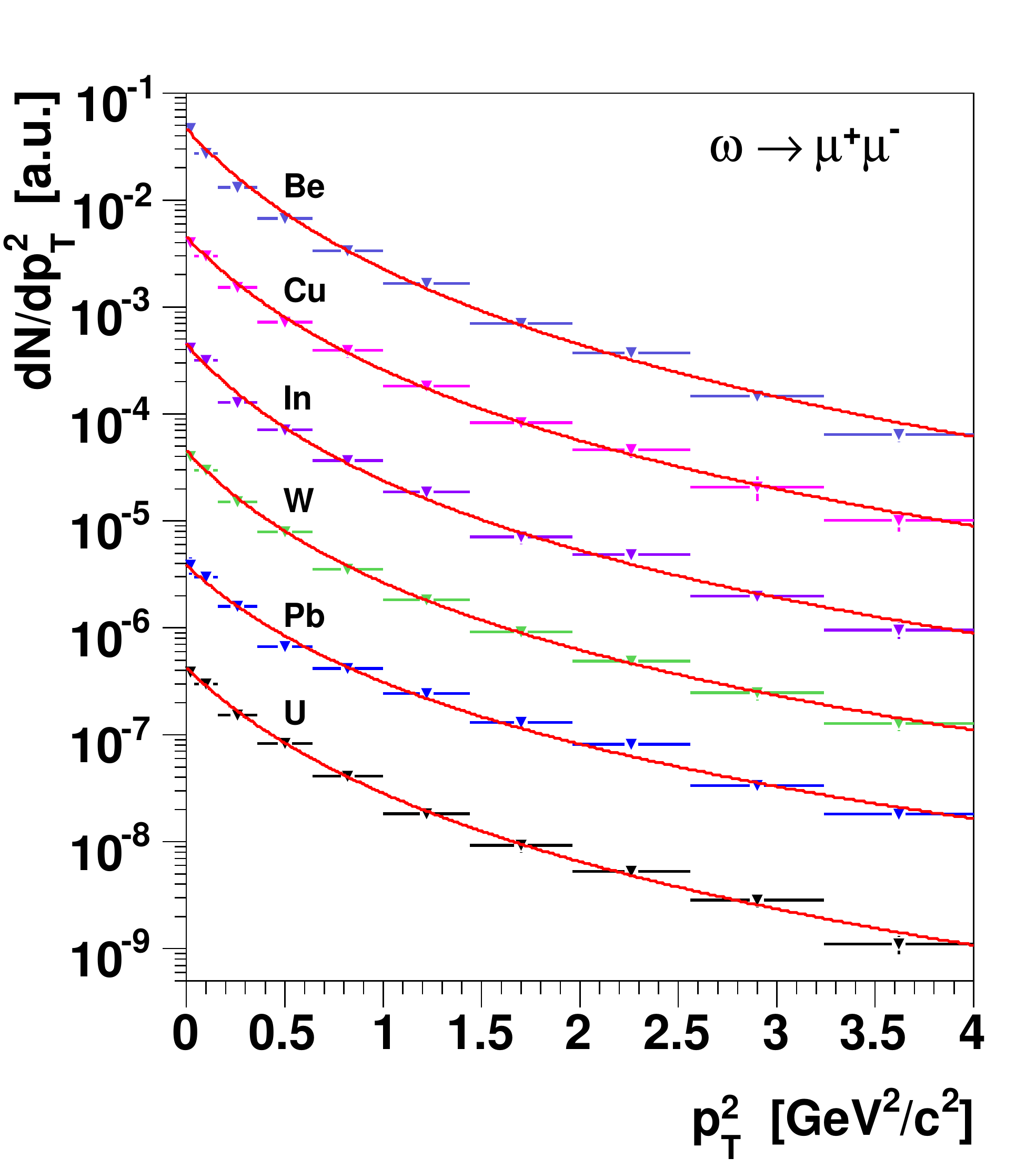}    \vspace*{-0.1cm}	

    \includegraphics[width=0.40\textwidth]{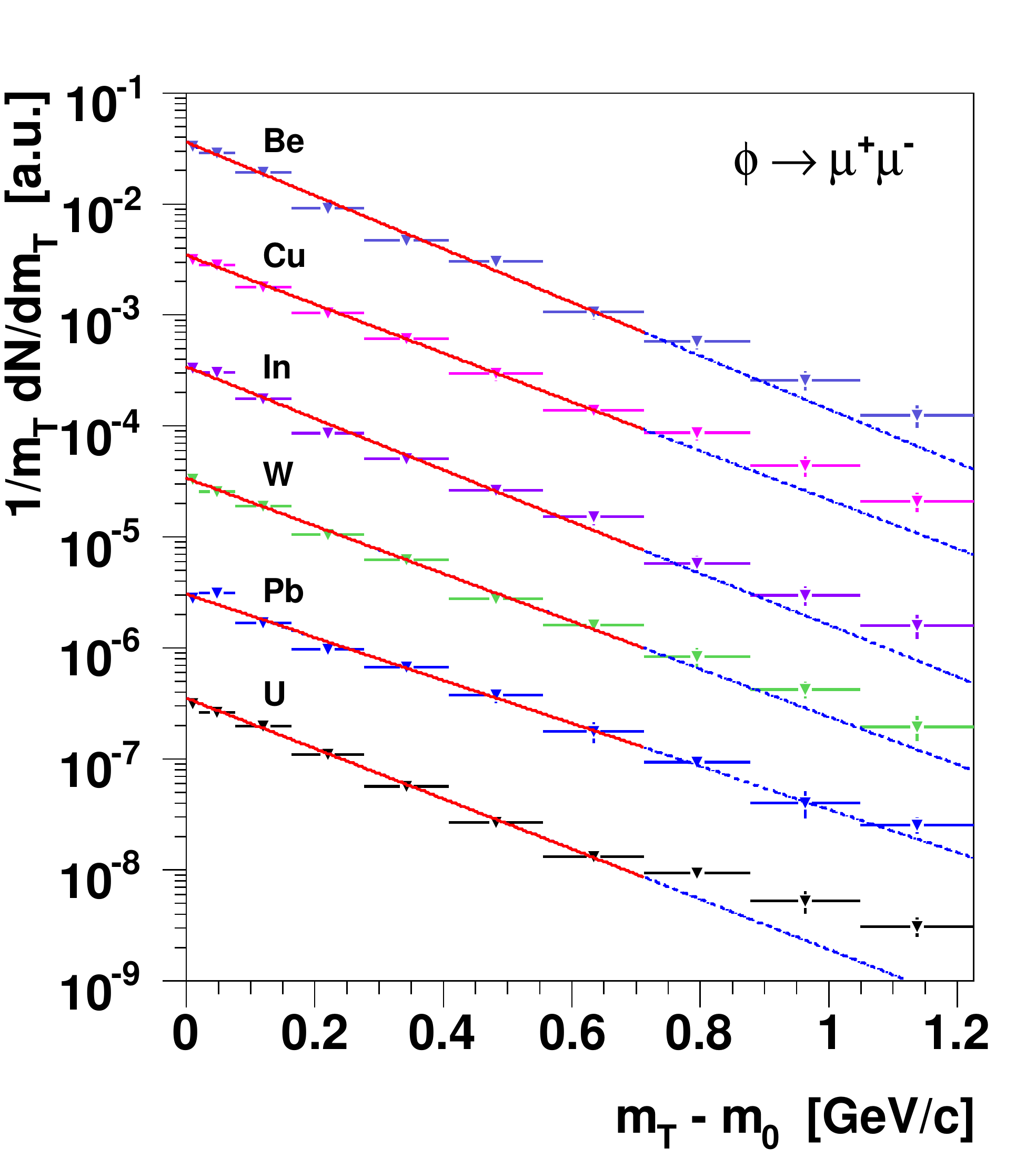}\hspace{0.05\textwidth}
    \includegraphics[width=0.40\textwidth]{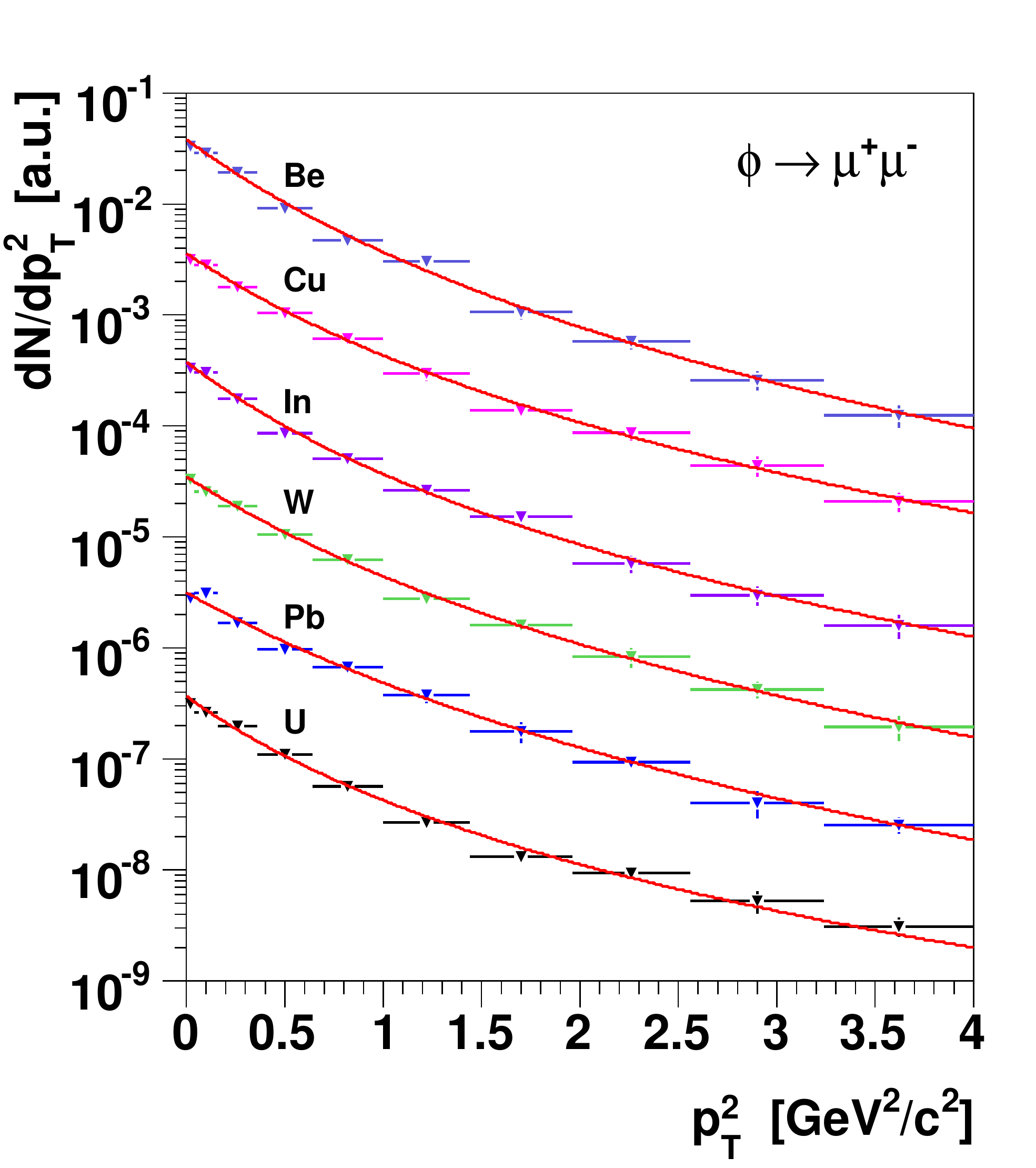}
  \end{center}
\vspace{-0.5cm}
\caption[\textwidth]{Fits on the acceptance-corrected $\mt$ and $\pt^2$ spectra}
\label{fig:pt_spectra}
\vspace{-0.3cm}
\end{figure}

~\\
\noindent In order to study the $\pt$ spectra of the $\omega$ and $\phi$ mesons, the data sample was divided, target by target, into $\pt$ bins of 200 MeV/$c$. For each target and $\pt$ bin, a fit was then performed on the raw mass spectrum with the superposition of the expected sources, using the same procedure exploited for the analysis of the target-integrated data sample (see left panel of \figurename~\ref{fig:mass_spectra}). In this way, a raw $\pt$ spectrum is extracted for the $\omega$ and $\phi$ mesons, which is then corrected for the acceptance $\times$ efficiency as a function of $\pt$, as given by the MC simulations. The $\omega$ and $\phi$ $\pt$ spectra resulting after the correction are shown in \figurename~\ref{fig:pt_spectra} as a function of the transverse mass $\mt$ and $\pt^2$. The $\mt$ spectra have been compared to a thermal-like function $ \d N/(\pt\, \d \pt) = \d N/(\mt \,\d \mt) \propto \exp ( -\mt/T)~$. As it can be seen, this thermal hypothesis systematically fails in describing the high $\mt$ tails deviating from the pure exponential trend. On the contrary, the power-law function $\d N/\d \pt^2 \propto \left( 1 + \pt^2/p_0^2 \right)^{-\beta}$ comes out to be adequate in describing the $\d N /\d \pt^2$ distributions over the whole range accessed by our data. No significant trend for the mean value $\langle \pt \rangle$ as function of the target size could be identified. The average $\pt$ integrated over the different nuclear targets are: $\langle \pt \rangle_\omega = 0.61 \pm 0.03~\mathrm{(stat.)} \pm 0.03~\mathrm{(syst.)}~\mathrm{GeV}/c$, $\langle \pt \rangle_\phi = 0.70 \pm 0.04~\mathrm{(stat.)} \pm 0.02~\mathrm{(syst.)}~\mathrm{GeV}/c$. These results can be compared with the NA27 measurements in p-p collisions at $\sqrt{s} = 27.5$~GeV~\cite{Verbeure:1991tv}: $\langle \pt \rangle_\omega^\mathrm{NA27} = (0.591 \pm 0.021)~\mathrm{GeV}/c$ and $\langle \pt \rangle_\phi^\mathrm{NA27} = (0.513 \pm 0.030)~\mathrm{GeV}/c$. As one can see, the NA27 value for $\langle \pt \rangle$ for the $\omega$ agrees with the one obtained in the present analysis. On the other hand, for the $\langle \pt \rangle$ of the $\phi$ there is a disagreement by more than 4 (statistical) standard deviations. Furthermore, the NA60 results clearly indicate that $\langle \pt \rangle_\phi > \langle \pt \rangle_\omega$, while the NA27 results indicate that $\langle \pt \rangle_\phi < \langle \pt \rangle_\omega$. Concerning the $\phi$ meson, HERA-B measured $\phi$ production in p-A at $\sqrt{s} = 41.6$~GeV~\cite{Abt:2006wt}. The $\pt$ distributions measured by HERA-B are well described by the power-law function cited above, with $\langle \pt \rangle$ values fully compatible with the NA60 ones and no definite trend as a function of the target size.

\begin{figure}[htbp]
   \begin{center} 
   \includegraphics[width=0.40\textwidth]{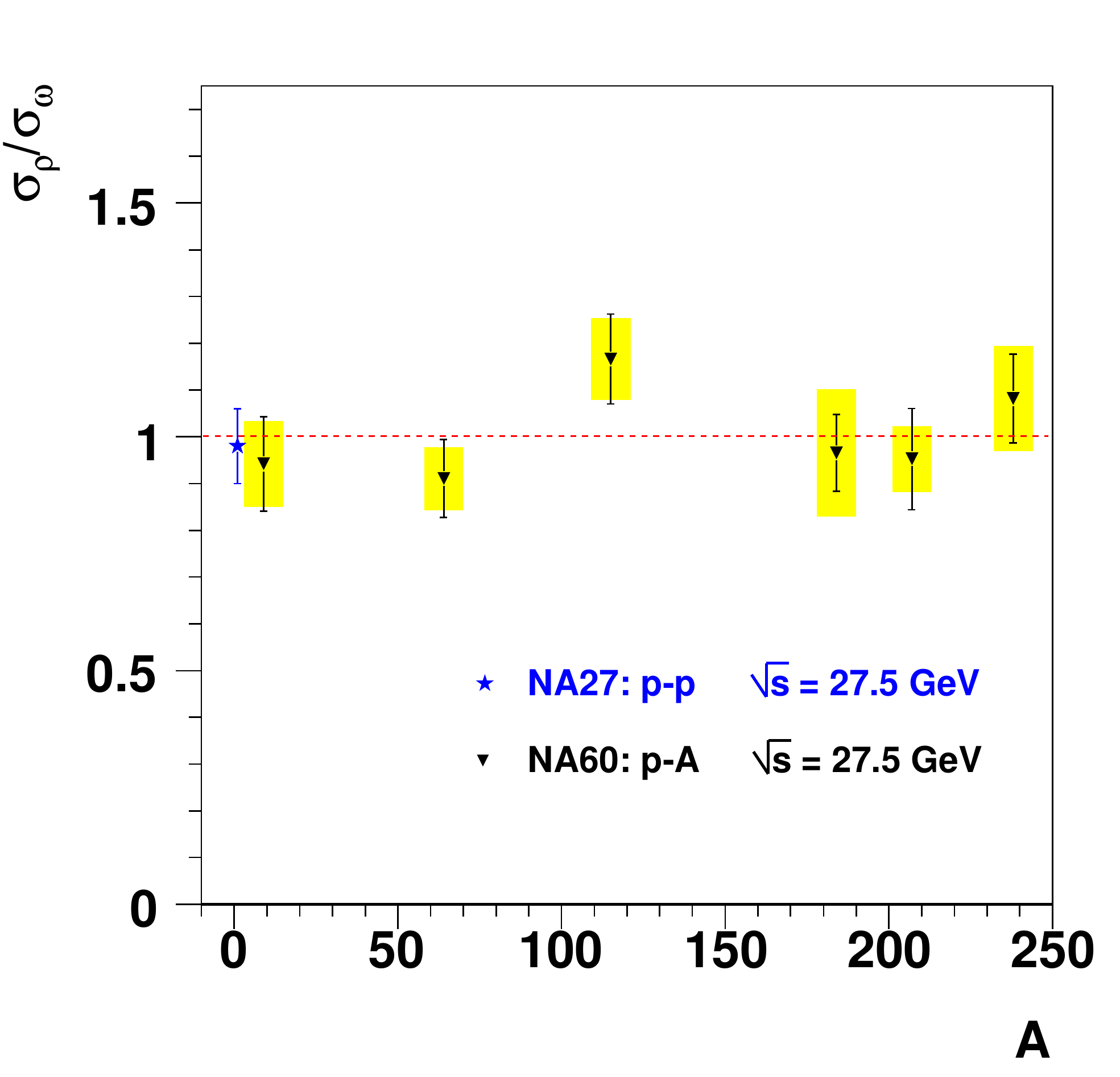} \hspace{0.10\textwidth} 
   \includegraphics[width=0.40\textwidth]{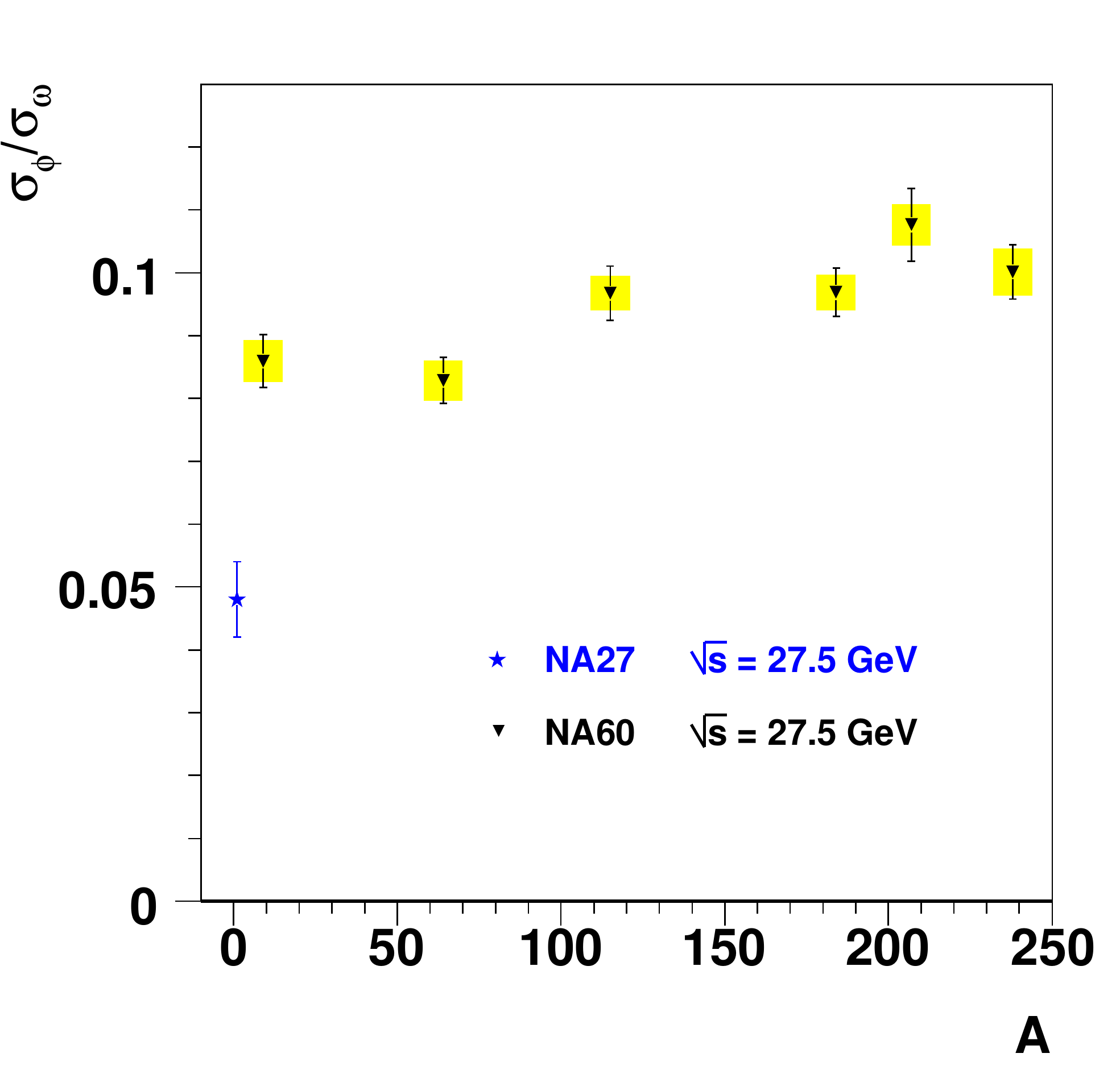}
    \end{center} 
\vspace{-0.7cm}
\caption[\textwidth]{Production cross section ratios $\sigma_\rho/\sigma_\omega$ and $\sigma_\phi/\sigma_\omega$ in full phase space.}
\label{fig:particleRatiosRhoPhi}
\vspace{-0.3cm}
\end{figure}

\noindent Fitting the raw mass spectra target by target allowed us to study the nuclear dependence of particle production. We first consider the particle ratios integrated over $\pt$, normalizing the cross sections to the one of the $\omega$ meson as shown in~\figurename~\ref{fig:particleRatiosRhoPhi}. The $\pt$-integrated yields and ratios considered here have been extracted for the $\rho$, $\omega$ and $\phi$ mesons only, since for the $\eta$ meson we have no acceptance for $\pt < 0.5$~GeV/$c$. The $\sigma_\rho/\sigma_\omega$ ratio appears to be flat with~A and consistent with $\sigma_\rho/\sigma_\omega = 1$. The ratio averaged over~A $-$ indicated by an horizontal line in the figure $-$ is $\sigma_\rho / \sigma_\omega = 1.00 \pm 0.04~\mathrm{(stat.)} \pm 0.04~\mathrm{(syst.)}$. This is in agreement with the ratio $\sigma_\rho/\sigma_\omega=0.98\pm0.08$ measured in p-p collisions at $\sqrt{s}=27.5$~GeV by the NA27 experiment~\cite{Verbeure:1991tv}. A rather different conclusion can be drawn, on the contrary, for the $\sigma_\phi/\sigma_\omega$ ratio, for which the p-p measurement by NA27~\cite{Verbeure:1991tv} is also shown: the trend of the NA60 data points suggests that a mild strangeness enhancement ($\sim 20\%$ from the lightest to the heaviest targets) is observed in p-A collisions as a function of the target size; on the other hand, the significant distance between the NA27 and the NA60 points makes it hard to reconcile the p-p and the p-A observations within a coherent picture.

\begin{figure}[htbp] 
   \begin{center}
    \includegraphics[width=0.32\textwidth]{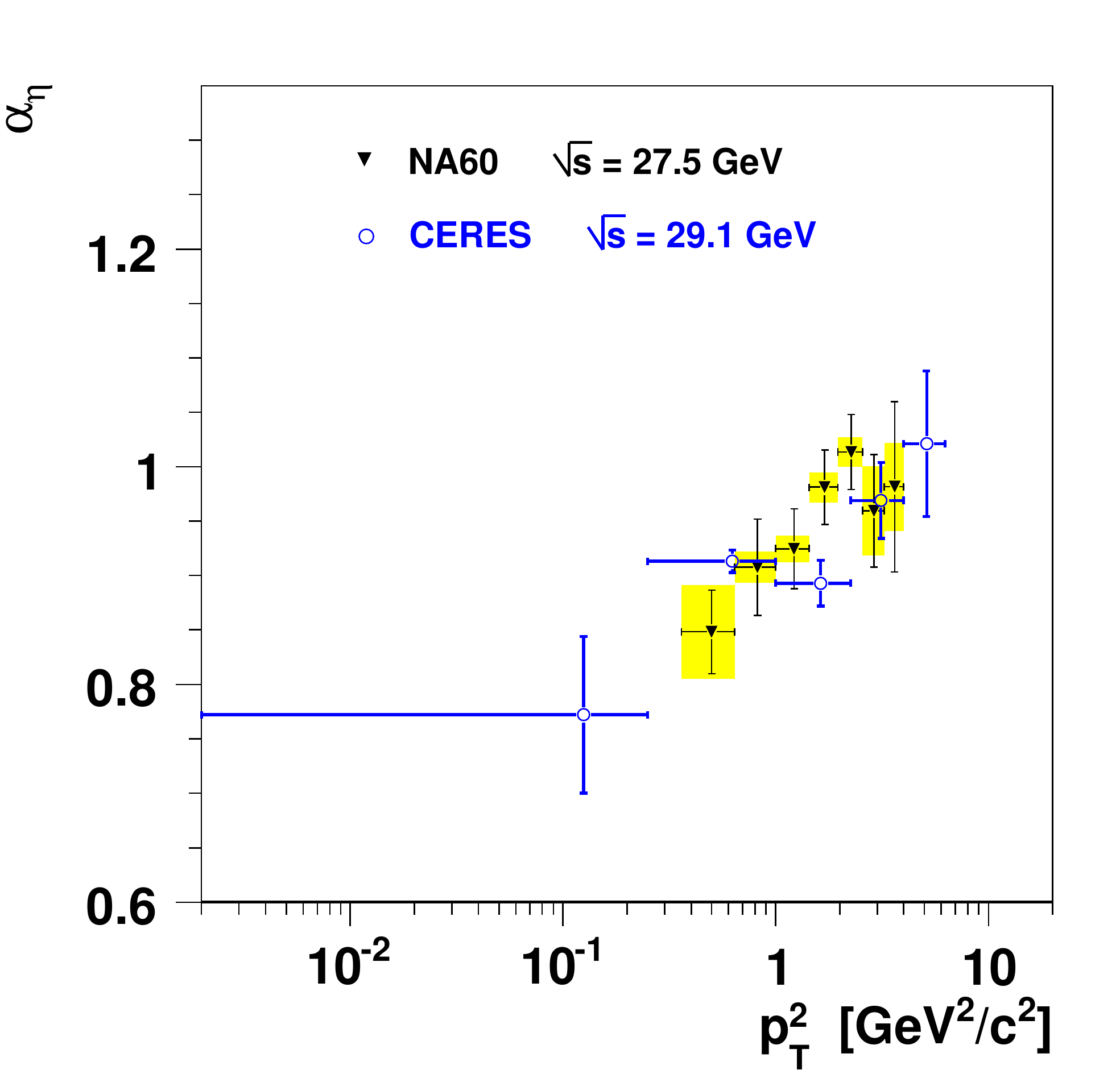}
    \includegraphics[width=0.32\textwidth]{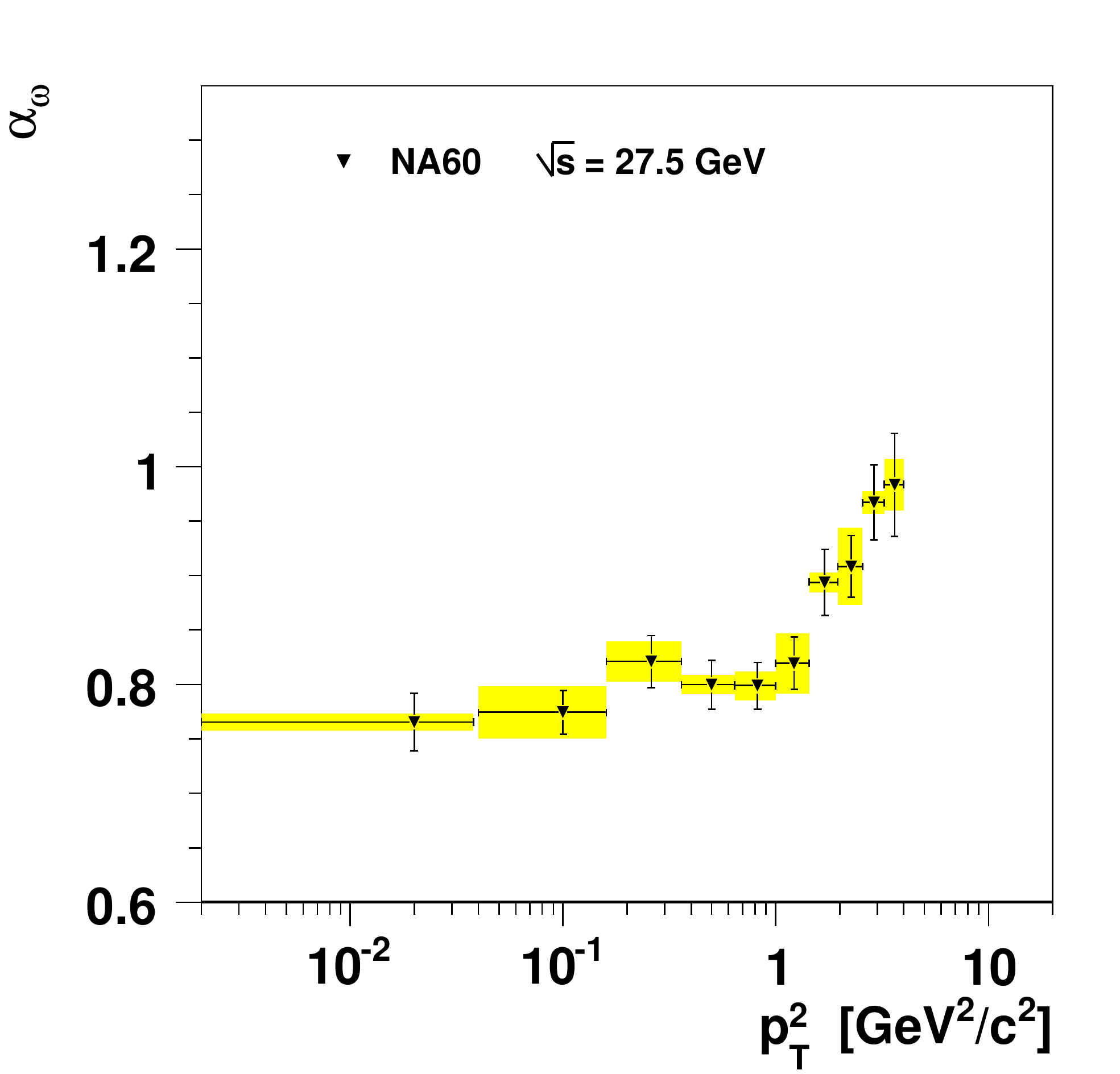} 
    \includegraphics[width=0.32\textwidth]{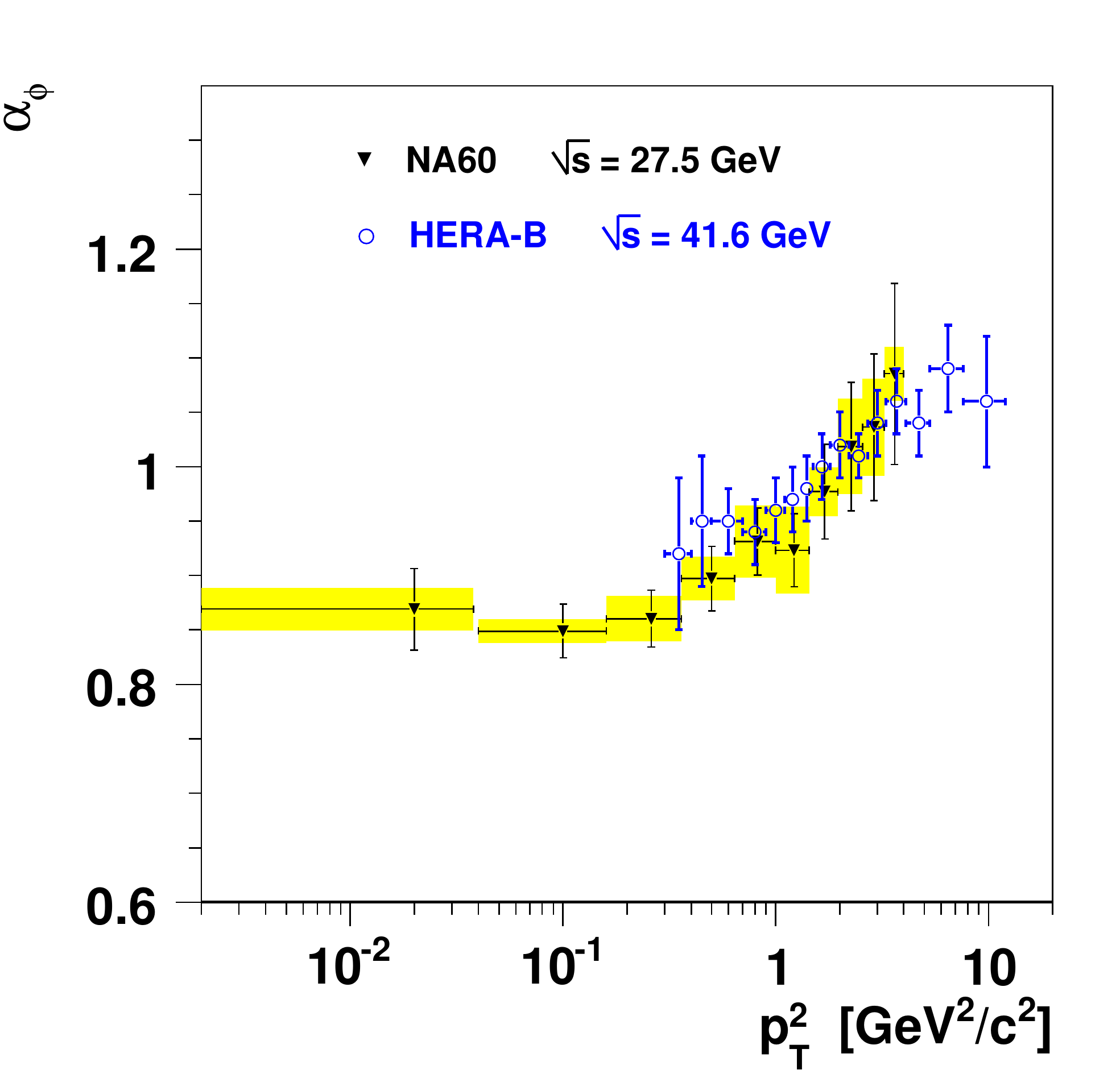}
    \end{center} 
\vspace{-0.5cm}
\caption[\textwidth]{$\pt$ dependence of the $\alpha$ parameter for the $\eta$, $\omega$ and $\phi$ mesons.}
\label{fig:alphaVsPt}
\end{figure}

~\\
\noindent The nuclear dependence of the production cross sections of the light mesons has been studied by normalizing the measured yields to the one of the lightest target, namely the Be one. The resulting trend is satisfactorily described by the power-law $\sigma_\mathrm{pA} \propto \mathrm{A}^\alpha$, leading to the following estimation for the $\alpha$ parameter for the $\omega$ and $\phi$ mesons, integrated over $\pt$: $\alpha_\omega = 0.841 \pm 0.014~\mathrm{(stat.)} \pm 0.030~\mathrm{(syst.)}$, $\alpha_\phi = 0.906 \pm 0.011~\mathrm{(stat.)} \pm 0.025~\mathrm{(syst.)}$.
These results confirm, from a different point of view, that the $\phi$ meson cross section rises with the target size faster than the one of the $\omega$ meson, coherently with the observation of a rising trend for the $\sigma_\phi/\sigma_\omega$ ratio with~A. The available data sample also allowed us to investigate the $\pt$ dependence of the $\alpha$ parameters, see \figurename~\ref{fig:alphaVsPt}. In this case we could directly compare the $\alpha$ parameters of the $\eta$ and the $\omega$ in the common range $\pt>0.6$~GeV/$c$: here, the average $\alpha$ parameter for the $\eta$ meson seems to be higher than the one of the $\omega$, suggesting a rising trend of the $\sigma_\eta/\sigma_\omega$ ratio with~A for $\pt > 0.6$~GeV/$c$. As a general results, the observed trends for the three mesons clearly indicate an increase of the $\alpha$ parameters as a function of $\pt$, related to the so-called ``Cronin effect'' originally observed by Cronin~\emph{et al.}~for charged kaons~\cite{Kluberg:1977bm}. The same observation was recently reported for the $\phi$ meson by the HERA-B Collaboration~\cite{Abt:2006wt}, whose points are in remarkable agreement with the NA60 ones. A similar comparison can be established for the $\eta$ meson, thanks to the data published by the CERES-TAPS Collaboration~\cite{Agakishiev:1998mw}: here, too, a good agreement is observed between the two sets of data points. No comparison was possible, conversely, for the $\omega$ meson, due to the lack of available measurements. 

\end{document}